\documentclass[twocolumn,aps, prb, showpacs,preprintnumbers,amsmath,amssymb,a4paper,superscriptaddress]{revtex4-1}

\usepackage{graphicx}
\usepackage{dcolumn}
\usepackage{bm}
\usepackage[percent]{overpic}

\begin{document}
\newcommand{\Arg}[1]{\mbox{Arg}\left[#1\right]}
\newcommand{\bb}{\mathbf}
\newcommand{\braopket}[3]{\left \langle #1\right| \hat #2 \left|#3 \right \rangle}
\newcommand{\braket}[2]{\langle #1|#2\rangle}
\newcommand{\be}{\[}
\newcommand{\br}{\vspace{4mm}}
\newcommand{\bra}[1]{\langle #1|}
\newcommand{\braketbraket}[4]{\langle #1|#2\rangle\langle #3|#4\rangle}
\newcommand{\braop}[2]{\langle #1| \hat #2}
\newcommand{\dd}[1]{ \! \! \!  \mbox{d}#1\ }
\newcommand{\DD}[2]{\frac{\! \! \! \mbox d}{\mbox d #1}#2}
\renewcommand{\det}[1]{\mbox{det}\left(#1\right)}
\newcommand{\ee}{\]} 
\newcommand{\eg}{\textbf{\\  Example: \ \ \ }}
\newcommand{\Imag}[1]{\mbox{Im}\left(#1\right)}
\newcommand{\ket}[1]{|#1\rangle}
\newcommand{\ketbra}[2]{|#1\rangle \langle #2|}
\newcommand{\kp}{\arccos(\frac{\omega - \epsilon}{2t})}
\newcommand{\ldos}{\mbox{L.D.O.S.}}
\renewcommand{\log}[1]{\mbox{log}\left(#1\right)}
\newcommand{\Log}{\mbox{log}}
\newcommand{\Modsq}[1]{\left| #1\right|^2}
\newcommand{\nb}{\textbf{Note: \ \ \ }}
\newcommand{\op}[1]{\hat {#1}}
\newcommand{\opket}[2]{\hat #1 | #2 \rangle}
\newcommand{\occ}{\mbox{Occ. Num.}}
\newcommand{\Real}[1]{\mbox{Re}\left(#1\right)}
\newcommand{\so}{\Rightarrow}
\newcommand{\sol}{\textbf{Solution: \ \ \ }}
\newcommand{\thetafn}[1]{\  \! \theta \left(#1\right)}
\newcommand{\tin}{\int_{-\infty}^{+\infty}\! \! \!\!\!\!\!}
\newcommand{\Tr}[1]{\mbox{Tr}\left(#1\right)}
\newcommand{\kb}{k_B}
\newcommand{\rad}{\mbox{ rad}}
\preprint{APS/123-QED}

\title{Strain-induced modulation of magnetic interactions in graphene}

\author{S.R. Power}
\email{spow@nanotech.dtu.dk}
\altaffiliation[Current affiliation: ]{Department of Micro- and Nanotechnology (DTU Nanotech), Technical University of Denmark, DK-2800 Kgs. Lyngby, Denmark.}
\affiliation{School of Physics, Trinity College Dublin, Dublin 2, Ireland}
\author{P. D. Gorman}
\affiliation{School of Physics, Trinity College Dublin, Dublin 2, Ireland}
\author{J. M. Duffy}
\affiliation{School of Physics, Trinity College Dublin, Dublin 2, Ireland}
\author{M. S. Ferreira}
\affiliation{School of Physics, Trinity College Dublin, Dublin 2, Ireland}
\affiliation{CRANN, Trinity College Dublin, Dublin 2, Ireland}

\date{\today}

\begin{abstract}
The ease with which the physical properties of graphene can be tuned suggests a wide range of possible applications. Recently, strain engineering of these properties has been of particular interest. Possible spintronic applications of magnetically-doped graphene systems have motivated recent theoretical investigations of the so-called Ruderman-Kittel-Kasuya-Yosida (RKKY) interaction between localized moments in graphene. In this work a combination of analytic and numerical techniques are used to examine the effects of uniaxial strain on such an interaction. A range of interesting features are uncovered depending on the separation and strain directions. Amplification, suppression and oscillatory behaviour are reported as a function of the strain and mathematically transparent expressions predicting these features are derived. 
Since a wide range of effects, including overall moment formation and magnetotransport response, are underpinned by such interactions we predict that the ability to manipulate the coupling by applying strain may lead to interesting spintronic applications.
\end{abstract}

                 
\maketitle

\section{Introduction}
Graphene has been attracting the attention of the wider scientific community due to an enormous range of tuneable properties, suggesting applications in fields as diverse as photonics, sensor technology and spintronics\cite{riseofgraphene, neto:graphrmp, yazyev:review}. 
In recent years, the potential to tune the electronic \cite{pereira_tight-binding_2009, Pereira09, Ribeiro09, Pereira10,Guinea:gapsgraphene, Ni08, Pellegrino11b, Neekamal12, PhysRevB.85.115432}, transport \cite{pereira_tight-binding_2009, Teague09, PhysRevB.81.161402, Bahat10, Pellegrino11, Klimov12, Kumar12}, optical \cite{Mohiuddin09, Pellegrino11b, Pellegrino10} and magnetic \cite{Levy30072010, Santos12, Santos12b, PhysRevB.84.075415, Peng20123434} properties of graphene systems by applying strain has been explored.
The degree to which these properties can be tuned is enhanced by the different types of strain that can be applied. Apart from simple uniaxial strains\cite{pereira_tight-binding_2009, Ni08}, more exotic features like creases and bubbles can be introduced \cite{Pereira10, Levy30072010, Klimov12, Neekamal12, Xu12, Neekamal12b}. 


An important topic in spintronics is the indirect exchange interaction between localized magnetic moments mediated by the conduction electrons of a conducting host medium. 
This interaction manifests itself as an energy difference between different alignments of the localized moments, leading to energetically favourable configurations.
Such an interaction is usually calculated within the Ruderman-Kittel-Kasuya-Yosida (RKKY) approximation\cite{RKKY:RK, RKKY:K, RKKY:Y} and indeed the interaction itself often takes this moniker. 
The RKKY interaction in graphene has been intensively studied \cite{Vozmediano:2005, PhysRevLett.97.226801, dugaev:rkkygraphene, saremi:graphenerkky, brey:graphenerkky, hwang:rkkygraphene, bunder:rkkygraphene, rapidcomm:emergence, black:graphenerkky, sherafati:graphenerkky, uchoa:rkkygraphene, black-schaffer_importance_2010, me:grapheneGF, sherafati:rkkygraphene2, kogan:rkkygraphene, disorderedRKKY, PhysRevB.84.205409, DynamicRKKY, Peng20123434} with a general consensus that the interaction strength decays asymptotically as $D^{-3}$ in undoped graphene\cite{obs}, where $D$ is the separation between the magnetic moments. 
This fast decay rate, arising from the graphene electronic structure at the Fermi energy, results in the interaction being very short-ranged. Any method of amplifying the coupling to extend its range is welcome and could prove useful for future spintronic applications. 
Another peculiar feature of this interaction in graphene-based materials is the masking of the usual sign-changing oscillations due to a commensurability effect\cite{AntonioDavidIEC}. 


With the motivation of amplifying the magnetic interaction strength, in this work we show how the magnetic coupling between localized moments in a graphene sheet can be manipulated by applying uniaxial strain. The sequence adopted is as follows. 
We start by introducing the general formalism used to calculate the magnetic coupling, which is written entirely in terms of the real-space single-particle Green functions (GF) of the host graphene sheet. 
We subsequently show in a mathematically transparent form how the GF is affected by the applied strain and use this result to predict the behaviour of the coupling when the direction or strength of the strain is varied. We find that both amplification and suppression of the magnetic coupling can be achieved. Furthermore we demonstrate that inter- and intra-sublattice couplings can be switched on and off independently, suggesting a wide range of possible applications. Our results are then confirmed using fully numerical calculations.


\section{Methods} 
\begin{figure}
 \centering
\includegraphics[width =0.4\textwidth]{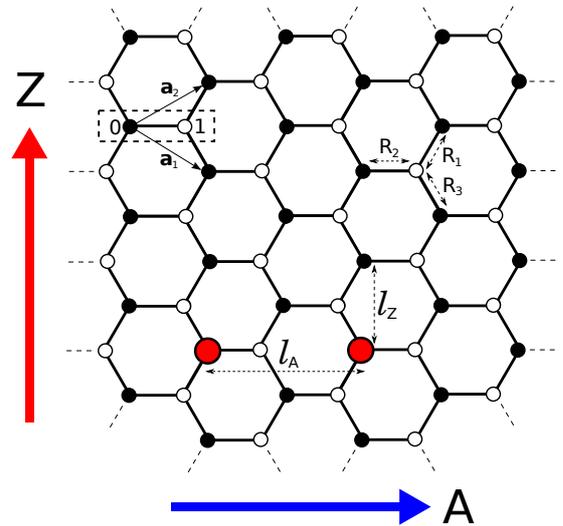}
\caption{Schematic representation of the graphene lattice with the armchair (zigzag) direction marked by the arrow labelled `A' (`Z'). The filled and hollow symbols represent sites on different sublattices. The unit vectors $\mathbf{a_1}$ and $\mathbf{a_2}$ and unit cell (dashed rectangle) are also shown. The large (red) symbols represent magnetic impurities in the lattice, in this case separated by the unit of separation in the armchair direction $D = l_A$. The unit of separation in the zigzag direction, $l_Z$, is also shown. The distances $R_1$, $R_2$ and $R_3$ between an atom in the lattice and its three nearest neighbours are shown.}
\label{fig-schematic}
\end{figure}

We start by considering two substitutional magnetic impurity atoms at sites $A$ and $B$ a distance $D$ apart embedded in a graphene sheet. Despite the simplicity of this setup, it is sufficient to capture the essence of the magnetic interaction between arbitrary magnetic objects. The indirect exchange coupling between these two moments can be calculated by considering the energy difference between the ferromagnetic (FM) and antiferromagnetic (AFM) alignments of the moments. The Lloyd formula method \cite{lloyd} can be employed to express this energy difference as 
\begin{equation}
J_{BA} = \frac{-1}{\pi} \,\mathrm{Im}  \int  \mathrm{d} E \, f(E) \, \ln  \left( 1 + 4 \, V_{ex}^2 \, \mathcal{G}_{BA}^{\uparrow} (E) \, \mathcal{G}_{AB}^{\downarrow} (E) \right) \,,
\label{lloyd_IEC}
\end{equation}
where $\mathcal{G}_{AB}^{\sigma}$ is the single-electron GF describing the propagation of electrons with spin $\sigma = \uparrow$ or $\downarrow$, $V_{ex}$ is the exchange splitting of the magnetic impurity and $f(E)$ is the Fermi function. To calculate the Green function we employ an Anderson-like Hamiltonian\cite{anderson_localized_1961} to describe the electronic properties of the system 
\begin{equation}
 \hat{H} =   \sum_{<\mathbf{r}, n, \mathbf{r}^\prime, n^\prime>, \sigma} t_{\mathbf{r},\mathbf{r}^\prime}^{n, n^\prime} \, \ {\hat c}_{\mathbf{r} n \sigma}^\dag \, {\hat c}_{\mathbf{r}^\prime n^\prime \sigma} + \sum_{\sigma, \alpha} \epsilon_\alpha^\sigma  \ {\hat c}_{\alpha\sigma}^\dag \, {\hat c}_{\alpha\sigma} \,.
 \label{hamiltonian}
\end{equation}
Here ${\hat c}_{\mathbf{r} n \sigma}^\dag$ (${\hat c}_{\mathbf{r} n \sigma}$) creates (annihilates) an electron with spin $\sigma$ in a $\pi$ orbital centred at site $n=0$ or $1$ in the two-atom unit cell shown in Fig. \ref{fig-schematic} whose location is given by $\mathbf{r}$. $t_{\mathbf{r},\mathbf{r}^\prime}^{n, n^\prime}$ is the electronic hopping term between two such orbitals. The sum in the first term is restricted to orbitals at neighbouring sites. Thus the first term in Eq. \eqref{hamiltonian} is the standard nearest-neighbour tight-binding Hamiltonian describing the graphene electronic bandstructure. The second term provides a simple description of the magnetic impurities at sites $\alpha = A, B$. The quantity $\epsilon_\alpha^\sigma = \pm V_{ex}$ is a spin-dependent onsite potential that accounts for the exchange splitting in the magnetic orbitals. In this model, we consider only a single magnetic orbital at each impurity site. It is straightforward to generalise this approach to deal with 
multiple orbitals and to model specific magnetic impurities more accurately by including additional terms in Eq. \eqref{hamiltonian} to modify, for example, the hopping between graphene and the impurity sites or the band centre of the impurities. Such a parameterisation is usually based on a comparison with \emph{ab initio} studies of single impurities embedded in graphene. A large number of studies of this kind have been performed for a wide range of possible magnetic objects\cite{arkady_embedding_2009, santos_first-principles_2010, zhang_electrically_2012, PhysRevX.1.021001, Hu12:5d}.  In this work we shall confine our discussion to generic impurities and the Hamiltonian above, since in general the properties of this type of interaction are determined by the host medium and are largely independent of the magnetic impurity species. 

The numerical results shown later in this paper use Eq. \eqref{lloyd_IEC} in conjunction with the Hamiltonian given in Eq. \eqref{hamiltonian}. However, to proceed analytically it is useful to note that Eq. \eqref{lloyd_IEC} can be written as a perturbation expansion in powers of the exchange splitting $V_{ex}$  and when expressed to leading order in $V_{ex}$ gives an expression equivalent to the commonly used RKKY approximation\cite{RKKY:Bruno2, Castro:Coupling}  
\begin{equation}
 J_{BA} = -\frac{4 \, V_{ex}^2}{\pi} \,  \int \, \mathrm{d} E \, f(E) \, \mathrm{Im} \left[\, G_{BA} (E) \, G_{AB} (E) \, \right] \,,
\label{static_J_rkky}
\end{equation}
where 
$G_{BA} (E)$ is the single-electron, spin-independent GF describing electron propagation in the pristine host material. In other words, the GF we consider corresponds to a Hamiltonian containing only the first term of Eq. \eqref{hamiltonian}. Each of the carbon orbitals considered in this model has three nearest neighbours, shown in Fig. \ref{fig-schematic}. In unstrained graphene, the bond lengths $R_1$, $R_2$ and $R_3$ are identical, and therefore so are the associated hopping terms $t_1$, $t_2$ and $t_3$ which take the value $t_0 = -2.7\,\mathrm{eV}$. When a tensile strain, $\varepsilon$, is applied to the graphene sheet the bond lengths and hence the hopping values are altered. We do not consider the effect of strain on the magnitude of the impurity moment. Several recent studies have discussed this topic for specific impurities in graphene\cite{Santos12, PhysRevB.84.075415, arkady_embedding_2009} and similar materials\cite{machado-charry:132405}. Ref. \onlinecite{PhysRevB.84.075415}, for example, 
considers the effects of strain on the magnetic moments of a wide range of transition metal impurities. In some cases, sudden jumps in the moment magnitude occur when the impurity changes its hybridisation with the graphene sheet at critical strain values. In other cases, for example substitutional Mn atoms, only a small change in the moment value is noted with increasing strain. Bearing in mind that the exchange splitting $V_{ex}$ is proportional to the moment value, Eq. \eqref{static_J_rkky} suggests that such changes in the moment value could influence the coupling, but should not qualitatively affect the results presented later in this work.

For uniaxial strains in the high symmetry zigzag (Z) and armchair (A) directions the bond lengths vary with strain as follows :
\begin{equation}
 \begin{aligned}
   Z\,: \, & \,  \tfrac{R_1}{R_0}  = \tfrac{R_3}{R_0}  = 1 + \tfrac{3}{4} \varepsilon - \tfrac{1}{4} \varepsilon \sigma \;, & \;\tfrac{R_2}{R_0} & = 1 - \varepsilon \sigma \\
A\,: \, & \,  \tfrac{R_1}{R_0}  = \tfrac{R_3}{R_0}  = 1 + \tfrac{1}{4} \varepsilon - \tfrac{3}{4} \varepsilon \sigma \;, & \;\tfrac{R_2}{R_0} & = 1 + \varepsilon \;,
 \end{aligned}
\label{new-bonds}
\end{equation}
where, $R_1=R_3$ due to symmetry, $R_0$ is the unstrained bond length in graphene and $\sigma=0.165$ is the graphite value for Poisson's ratio, giving the level of contraction in the direction perpendicular to the applied strain.
The hopping parameters vary with the change in bond length, $\Delta R$ as
\begin{equation}
 t (\Delta R) = t_0 \, e^{-\alpha \frac{\Delta R}{R_0}} \,,
\label{new-hoppings}
\end{equation}
where $\alpha =3.37$ is taken from the literature \cite{pereira_tight-binding_2009, PhysRevB.75.045404}. 
Using the hopping parameters found using  Eqs. \eqref{new-bonds} and \eqref{new-hoppings} allows us to calculate the bandstructure of a strained graphene system. The dispersion relation \cite{pereira_tight-binding_2009}, is given by 
$
 \epsilon_{\pm} = \pm \sqrt{t_2^2 + 4\, t_1 t_2 \, \cos k_A \cos k_Z + 4 \, t_1^2 \, \cos^2 k_Z} \;.
$
For convenience we have defined dimensionless $k$-space vectors 
$
k_A = \tfrac{1}{2} \, l_A \, k_x \quad, \;k_Z = \tfrac{1}{2} \, l_Z \, k_y
$
in terms of $l_A$ ($l_Z$) - the strained unit of length between unit cells separated in the armchair (zigzag) direction and shown in Fig. \ref{fig-schematic}, where $k_x$ and $k_y$ are the reciprocal vectors in the $x$ and $y$ directions. It is important to distinguish between the strain and separation directions. Both armchair and zigzag separations between the moments will be considered, and for both cases strains will be applied parallel and perpendicular to the separation direction. In our convention, the zigzag and armchair directions are mutually perpendicular so that strain applied parallel (perpendicular) to the separation direction is applied along the same (opposite) high symmetry direction. 

The real-space GF between two sites on the graphene lattice separated by a vector $\mathbf{D}$  can be written as a double integral over the Brillouin Zone. We have shown previously\cite{me:grapheneGF} for unstrained graphene that one of the integrals can be solved analytically using contour integration and that for high-symmetry direction separations, the remaining integral is very well approximated using the Stationary Phase Approximation. This approach allows us to write the GF for energies throughout the entire band in the form
 \begin{equation}
{\cal G}_{D}(E) = {{\cal A}(E) \, e^{i \mathcal{Q} (E) D} \over \sqrt{D}} \,,
\label{conciseGF}
\end{equation}
where ${\cal A}(E)$ is an energy-dependent coefficient and $\mathcal{Q} (E)$ can be identified with the Fermi wavevector in the direction of separation. The exact functional forms of these quantities depend on the separation direction, but the distance dependence of the GF is clear in this form. Following Ref. \cite{me:grapheneGF} we can generalise the expressions to strained graphene.

%

\begin{figure}
 \centering
\includegraphics[width =0.4\textwidth]{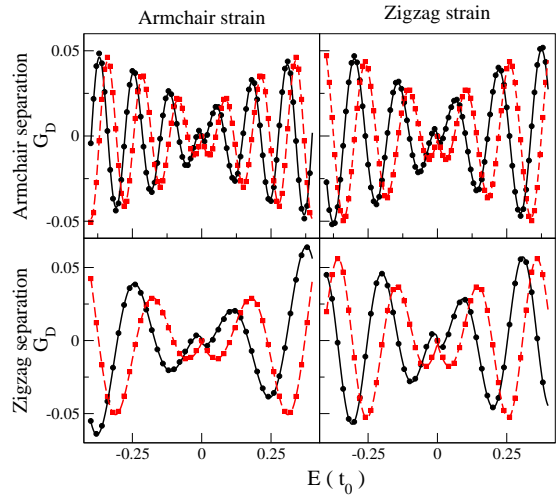}
\caption{Comparison between numerically calculated (symbols) and analytic expressions (lines) for the GF between two sites on the same sublattice in strained graphene systems. In all cases black solid lines and circles (red dashed lines and squares) represent the real (imaginary) part of the GF. The upper panels represent the GF for a separation of 40 $l_A$ in the armchair direction and the lower panels a separation of $40 l_Z$ in the zigzag direction. The left (right) panel in each case represents the GF for a strain in the armchair (zigzag) direction of $\varepsilon = 0.05$. An excellent agreement is seen for each.}
\label{fig-gf}
\end{figure}

For separations in the armchair direction between sites on the same sublattice, and for energy values in a broad range around $E=0$ ($|E| \lesssim 0.5 |t_0| $), we can write
\begin{equation}
 \begin{split}
  {\cal A}(E, \varepsilon) & = \sqrt{\frac{2}{i\pi}} \sqrt{ \frac{- E}{(E^2 + 4 t_1^2 - t_2^2) \sqrt{t_2^2 - E^2}}} \\
  {\cal Q}(E, \varepsilon) & = \cos^{-1} \left(\frac{\sqrt{t_2^2 - E^2}}{t_2} \right) \,.
 \end{split}
\label{acAQ}
\end{equation}
For zigzag separations there are two contributions ($\pm$) to the GF, whose corresponding expressions are
\begin{equation}
 \begin{split}
  {\cal A}_\pm(E, \varepsilon) & =  \sqrt{\frac{1}{2 i \pi }} \;  \sqrt{\frac{E}{|t_2| (t_2 \pm E)}} \; \frac{1}{\left( 4 t_1^{\,2} - (E \pm t_2)^{\,2}\;\right)^{\;1/4}}\\
  {\cal Q}_\pm(E, \varepsilon) & = \cos^{-1} \left(\frac{- t_2 \mp E}{2t_1} \right) \,.
\end{split}
\label{zzAQ}
\end{equation}
The strain dependence in these cases enters through $t_1$ and $t_2$, given by Eqs. \eqref{new-bonds} and \eqref{new-hoppings}. In Fig \ref{fig-gf} we demonstrate the remarkable agreement between these expressions and numerically calculated GFs for a representative sample of separation and strain directions. The analytic form of these expressions should prove useful since many physical properties can be written in terms of Green function elements. Furthermore they are not limited solely to the linear dispersion regime and are valid across a large energy range.

\section{RKKY interaction in strained graphene} 
The behaviour of the magnetic coupling can be extracted from Eq. \eqref{static_J_rkky} quite easily when the GFs are expressed in the form shown in  Eq. \eqref{conciseGF}. The integration procedure is identical to that for unstrained graphene \cite{me:grapheneGF} and can be reduced to a sum over Matsubara frequencies. The functions ${\cal B}(E, \varepsilon) = {\cal A}^2(E, \varepsilon)$ and ${\cal Q}(E, \varepsilon)$ are expanded around $E_F$ and in the low temperature limit $T \rightarrow 0$, we find 

\begin{equation}
J_{BA}  \sim \mathrm{Im} \,\sum_{\ell} \frac{\mathcal{J}_\ell (E_F ,\varepsilon) }{D^{\,\ell+2}} \, \cos\,(2  \mathcal{Q}(E_F, \varepsilon)\, D)
\label{spa_coupling}
\end{equation}
where
\begin{equation}
 \qquad \mathcal{J}_\ell (E_F ,\varepsilon) = \frac{(-1)^\ell \, V_{ex}^2 \, \mathcal{B}^{(\ell)} (E_F ,\varepsilon) }{(2 \mathcal{Q}^\prime (E_F ,\varepsilon))^{\,\ell+1} }
\label{coupling_coeff}
\end{equation}
is the distance-independent coefficient for the $\ell$-th term in the power series, $\ell$ is a non-negative integer and $\mathcal{B}^{(\ell)} (E_F ,\varepsilon)$ is the $\ell$-th order energy derivative of $\mathcal{B} (E ,\varepsilon)$ evaluated at $E_F$, resulting from its Taylor expansion. In general the leading term in the series should determine the asymptotic decay rate of the coupling. For the undoped case it can be shown, for both strained and unstrained graphene, using Eqs. \eqref{acAQ} and \eqref{zzAQ} that the coefficient $\mathcal{B}^{(0)} (0 ,\varepsilon) =0$, so that the $\ell=1$ term dominates and $J(E_F=0) \sim D^{-3}$. Thus we should not expect to change the decay rate of the interaction by applying uniaxial strain. To study how strain does affect the coupling, we examine Eq. \eqref{spa_coupling} in undoped graphene as strain is applied and then increased. We define the strain-dependent amplification, $\beta$, as the ratio between the strained and unstrained couplings,
\begin{equation}
 \beta (\varepsilon) = \frac{J_{BA} (\varepsilon)}{ J_{BA} (\varepsilon=0)} \,.
\label{amplitude_factor}
\end{equation}

\subsection{Armchair separations} 
\begin{figure}
 \centering
\includegraphics[width =0.45\textwidth]{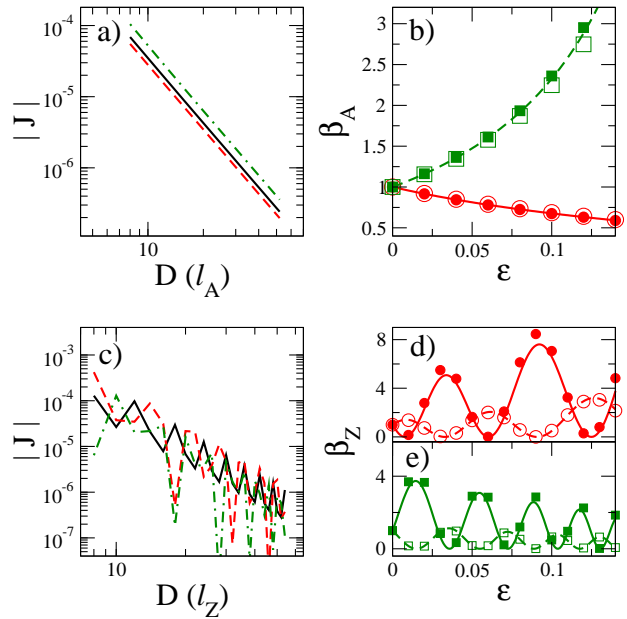}
\caption{a) Log-log plot showing the numerical magnetic coupling against armchair-direction separation without strain (black, solid line) and with $\varepsilon=0.05$ armchair (red, dashed) and zigzag (green, dot-dashed) uniaxial strain applied. b) Amplification factor $\beta_A$ as a function of strain in the armchair (red, solid line) and zigzag (green, dashed) directions. Filled and hollow symbols represent numerical calculations for same-sublattice and opposite sublattice cases respectively. c) Same as a) but for zigzag separation of impurities. d) Amplification factor $\beta_Z$ as a function of armchair strain for same (solid line, filled symbols) and opposite (dashed line, hollow symbols) sublattice cases. Lines represent the analytic result and symbols the numerical calculations. e) Same as d) but for zigzag strain.}
\label{fig-coupling}
\end{figure}

The periodicity of the coupling, determined from ${\cal Q}(E, \varepsilon)$ in Eq. \eqref{acAQ}, is clearly independent of $t_1$ and $t_2$, and thus strain, for $E=0$. Thus the only effect that strain can have is a distance-independent amplification or suppression arising from the $\mathcal{J}_1$ term in Eq. \eqref{spa_coupling}. In Fig \ref{fig-coupling}a) we plot the numerically-calculated coupling between moments on the same sublattice as a function of armchair separation. This quantity is shown for zero strain (black line), and for strains of $\varepsilon=0.05$ in the armchair (red, dashed) and zigzag (green, dot-dashed) directions. The results, shown in log-log form, confirm that the decay rate is unaffected as the strained cases lie parallel to the unstrained case. We note that the coupling is enhanced by zigzag strain, and suppressed by armchair strain. To study the effect of increasing strain we calculate the amplification factor $\beta(\varepsilon)$. Using Eq. \
eqref{acAQ}  we find a simple analytical form for armchair separations
\begin{equation}
 \beta_A = 3 \,t_0 \; \frac{t_2}{4 t_1^2 - t_2^2} \,.
\label{beta_A}
\end{equation}
We note that this expression has the same form in terms of $t_1$ and $t_2$ for strains in both high symmetry directions, but that the $t_1$ and $t_2$ values themselves depend on the strain direction, as given by Eq. \eqref{new-bonds}. In Fig \ref{fig-coupling}b) we plot the analytic expression for $\beta_A$ as a function of strain in both the armchair (red, solid line) and zigzag (green, dashed line) strain directions. A monotonic decrease (increase) in the coupling for armchair (zigzag) separations consistent with the results in panel a) is observed. To confirm the analytic prediction, numerical calculations of $\beta_A$ are performed for a fixed value of separation ($D = 20\, l_A$). Filled and hollow symbols represent calculations performed for sites on the same or different sublattice(s) respectively. An excellent agreement with the analytic predictions is seen in all cases. In addition to the substitutional case discussed here, numerical calculations were also performed for the case of impurities 
adsorbed on top of a single carbon atom. The results (not shown) are also in perfect agreement with Eq. \eqref{beta_A}, highlighting the fact that an exact parameterisation of the magnetic impurity is not necessary to calculate the qualitative behaviour of the RKKY interaction.  The identical amplification of couplings between same and opposite sublattice sites is explained by the phase factor between these couplings, which is zero for armchair direction separations \cite{sherafati:graphenerkky}. For other separation directions a more complex behaviour is expected as this phase factor is no longer zero, and the phase of the distant-dependent oscillations may also be strain dependent.

\subsection{Zigzag separations}
In Fig \ref{fig-coupling}c) we show the magnetic coupling as a function of zigzag separation for the unstrained case (black, solid line) and for strains of $\varepsilon=0.05$ in the armchair (red, dashed) and zigzag (green, dot-dashed) directions. A more complicated, non-monotonic behaviour than the armchair case is observed. This arises due to the Fermi wavevector in the zigzag direction and for unstrained graphene the oscillation has a period of $3 l_Z$.  As strain is applied, the wavevector determining the oscillation period varies as
$
{\cal Q}(\varepsilon) = {\cal Q}(0) \, + \, \delta k(\varepsilon)
$
where
$
 \delta k = \cos^{-1}\left(\frac{-t_2}{2t_1}\right) - \frac{4\pi}{3}
$. The amplification factor in the zigzag direction is thus 
\begin{equation}
 \beta_Z = \frac{t_0 \, \sqrt{4 t_1^2 - t_2^2}}{\sqrt{3}\, t_2^2} \,  \, \frac{\cos^2 \left(\,(\mathcal{Q}(0) + \delta k (\varepsilon) )\, D \right)}{\cos^2(\mathcal{Q}(0)\, D)} \,.
\label{beta_Z}
\end{equation}
The first part is a distance-independent term similar to $\beta_A$ which gives a monotonic increase (decrease) in the coupling for strain applied in the armchair (zigzag) direction. Viewed with the armchair results, this suggests a trend of strain perpendicular (parallel) to the separation direction amplifying (suppressing) the coupling. The second part of Eq. \eqref{beta_Z} accounts for amplification due to the change in the Fermi wavevector with strain and leads to oscillations in $\beta_Z$. The analytic expression for $\beta_Z$ is plotted in Fig. \ref{fig-coupling}d) and e) for armchair and zigzag strains respectively for both same-sublattice (solid lines) and opposite-sublattice (dashed line) cases with $D=40 l_Z$. The opposite-sublattice results take into account the $\frac{\pi}{2}$ phase shift from the same-sublattice case predicted for zigzag separations \cite{sherafati:graphenerkky}. An excellent agreement is again noted with the numerical calculations represented by filled (same-sublattice) and 
hollow (opposite-sublattice) symbols. The oscillations in the coupling, which appear as a function of strain, are very interesting and may have significant implications for strain-tuning of the interaction. Unlike for armchair separations, a small difference in the applied strain can tune the coupling from zero to several multiples of the unstrained value. Since the same-sublattice and opposite-sublattice couplings are exactly out of phase in this direction one is switched off when the other reaches a maximum. For an arbitrary non-armchair separation the coupling will have characteristic strain values for which one of the couplings is zero but the other is not. Thus strain suggests itself as a powerful tool, not only to amplify the interaction between impurity moments, but also to switch the interaction on and off and to control the interplay between impurities on different sublattices.

\section{Conclusions}
In this work we have derived analytic expressions for the Green function and RKKY interaction in graphene for high symmetry directions (armchair and zigzag) of both separation and applied uniaxial strain. Since GF methods are used to describe a wide range of physical properties, our expressions should prove useful in the investigation of strained graphene systems. An excellent match is found between these analytical expressions and full numerical calculations. 
Similarly, the simple closed-form expressions describing the amplification of the magnetic interaction in a strained graphene system agree with our numerical results.
A general trend of amplification for strain perpendicular to the moment separation direction, or supression for strain parallel to this direction, is noted. Also noted are oscillations in the amplification as strain is increased for moments separated in a non-armchair direction. This behaviour is again well-captured by our analytic approach. 
Such oscillations suggest the intriguing possibility of selectively turning on or off the coupling between moments and in particular of controlling the inter- and intra-sublattice couplings independently. Since the magnetic coupling underpins a wide range of physical features, including overall moment formation and magnetotransport response, the ability to fine tune the coupling with strain may lead to interesting spintronic applications. We hope that further investigation of strained graphene systems with magnetic impurities will yield a diverse range of tuneable properties suitable for device application. Finally, we would like to note that we recently became of a similar work by Peng and Hongbin\cite{Peng20123434}.

\begin{acknowledgments}
The authors acknowledge financial support received from the Irish Research Council for Science, Engineering and Technology under the EMBARK initiative and from Science Foundation Ireland under Grant Number SFI 11/RFP.1/MTR/3083. Computational resources were provided on the Lonsdale cluster maintained by the Trinity Centre for High Performance Computing. This cluster was funded through grants from Science Foundation Ireland. 
\end{acknowledgments}


\begin{thebibliography}{61}
\bibitem [{obs()}]{obs}%
  \BibitemOpen
  \href@noop {} {}\bibinfo {note} {Undoped graphene represents graphene with
  its Fermi level at half filling.}\BibitemShut {Stop}%
\end{thebibliography}
\end{document}